\documentclass[letterpaper]{article} % DO NOT CHANGE THIS
\usepackage{aaai23}  % DO NOT CHANGE THIS
\usepackage{times}  % DO NOT CHANGE THIS
\usepackage{helvet}  % DO NOT CHANGE THIS
\usepackage{courier}  % DO NOT CHANGE THIS
\usepackage[hyphens]{url}  % DO NOT CHANGE THIS
\usepackage{graphicx} % DO NOT CHANGE THIS
\usepackage{natbib}  % DO NOT CHANGE THIS AND DO NOT ADD ANY OPTIONS TO IT
\usepackage{caption} % DO NOT CHANGE THIS AND DO NOT ADD ANY OPTIONS TO IT
\usepackage{physics, amsmath}
\usepackage{multirow}
\usepackage[caption=false]{subfig}
\usepackage{comment}
\usepackage{booktabs}
\usepackage{amsfonts}

\DeclareMathOperator*{\argmin}{arg\,min}
\usepackage{algorithm}
\usepackage{algorithmic}

\usepackage{newfloat}
\usepackage{listings}
\DeclareCaptionStyle{ruled}{labelfont=normalfont,labelsep=colon,strut=off} % DO NOT CHANGE THIS
\floatstyle{ruled}
\newfloat{listing}{tb}{lst}{}
\floatname{listing}{Listing}
\title{A Vector Quantized Approach for Text to Speech Synthesis on Real-World Spontaneous Speech}
\author{
    Li-Wei Chen,
    Shinji Watanabe,
    Alexander Rudnicky
}
\affiliations{
    Language Technologies Institute, Carnegie Mellon University\\
    liweiche@cs.cmu.edu, swatanab@cs.cmu.edu, air@cs.cmu.edu
}

\begin{document}

\maketitle

\begin{abstract}
  Recent Text-to-Speech (TTS) systems trained on reading or acted corpora have achieved near human-level naturalness.
  The diversity of human speech, however, often goes beyond the coverage of these corpora.
  We believe the ability to handle such diversity is crucial for AI systems to achieve human-level communication.
  Our work explores the use of more abundant real-world data for building speech synthesizers.
  We train TTS systems using real-world speech from YouTube and podcasts.
  We observe the mismatch between training and inference alignments in mel-spectrogram based autoregressive models, leading to unintelligible synthesis, and demonstrate that learned discrete codes within multiple code groups effectively resolves this issue.
  We introduce our MQTTS system whose architecture is designed for multiple code generation and monotonic alignment, along with the use of a clean silence prompt to improve synthesis quality.
  We conduct ablation analyses to identify the efficacy of our methods.
  We show that MQTTS outperforms existing TTS systems in several objective and subjective measures.
\end{abstract}

\section{Introduction}
\label{sec:intro}
A crucial component of Artificial Intelligence (AI), especially conversational agents, is the ability to synthesize human-level speech.
With recent advances in deep learning, neural-based Text-to-Speech (TTS) systems~\cite{Li_Liu_Liu_Zhao_Liu_2019,NEURIPS2020_5c3b99e8,NEURIPS2019_f63f65b5} have led to significant improvements in the quality of synthesized speech.
However, standard corpora~\cite{ljspeech17,VCTK} used for training TTS systems for the most part include reading or acted speech recorded in a controlled environment.
On the other hand, humans spontaneously produce speech with diverse prosody that conveys paralinguistic information including subtle emotions. This ability comes from exposure to thousands of hours of real-world speech.
It suggests that TTS systems trained on real-world speech enable human-level AI.

In this work, we explore the use of real-world speech collected from YouTube and podcasts on TTS.
While the ultimate goal is to use an ASR system to transcribe real-world speech, here we simplify the setting by using an already transcribed corpus and focus on TTS.
Systems successfully trained on real-world speech are able to make use of the unlimited utterances in the wild.
Therefore, we believe it should be possible to replicate the success similar to that of large language models (LLMs) such as GPT-3~\cite{NEURIPS2020_1457c0d6}.
These systems can be fine-tuned to specific speaker characteristics or recording environments with few resources available.
In this work, we address emerging challenges when training TTS systems on real-world speech: higher variation of prosody and background noise compared to reading speech recorded in controlled environments.

With real-world speech, we first provide evidence that mel-spectrogram based autoregressive systems failed to generate proper text-audio alignment during inference, resulting in unintelligible speech.
We further show that clear alignments can still be learned in training, and thus the failure of inference alignment can be reasonably attributed to error accumulation in the decoding procedure.
We find that replacing mel-spectrogram with learned discrete codebooks effectively addressed this issue.
We attribute this to the higher resiliency to input noise of discrete representations.
Our results, however, indicate that a single codebook leads to distorted reconstruction for real-world speech, even with larger codebook sizes.
We conjecture that a single codebook is insufficient to cover the diverse prosody patterns presented in spontaneous speech.
Therefore, we adopt multiple codebooks and design specific architectures for multi-code sampling and monotonic alignment.
Finally, we use a clean silence audio prompt during inference to encourage the model on generating clean speech despite training on a noisy corpus.
We designate this system MQTTS (multi-codebook vector quantized TTS) and introduce it in Section~\ref{sec:model}.

We perform ablation analysis in Section~\ref{sec:res}, as well as comparing mel-spectrogram based systems to identify the properties needed for real-world speech synthesis.
We further compare MQTTS with non-autoregressive approach.
We show that our autoregressive MQTTS performs better in intelligibility and speaker transferability.
MQTTS achieves slightly higher naturalness, and with a much higher prosody diversity.
On the other hand, non-autoregressive model excels in robustness and computation speed.
Additionally, with clean silence prompt, MQTTS can achieve a much lower signal-to-noise ratio (SNR).
We make our code public\footnote{https://github.com/b04901014/MQTTS}.%, and hope that this work can bring us one step closer to human-level conversational agents.

\section{Related Work}

\textbf{Autoregressive TTS.}
Typical autoregressive TTS systems use a mel-spectrogram--vocoder pipeline~\cite{8461368,Li_Liu_Liu_Zhao_Liu_2019}.
An autoregressive model is trained to synthesize a mel-spectrogram sequentially, then a neural vocoder is separately trained  to reconstruct the waveform.
One concern in these systems is their low tolerance for variance  in the training corpus.
For instance, the alignment learning of Tacotron 2~\cite{8461368} is sensitive even to leading and trailing silences.
Such characteristics make training on highly noisy real-world speech infeasible.
We show empirically that this is resolved by replacing mel-spectra with quantized discrete representation.
Another recurring issue is faulty alignments during inference, leading to repetition and deletion errors.
Research has shown that this can be mitigated by enforcing monotonic alignment~\cite{DBLP:conf/interspeech/HeDH19,DBLP:journals/corr/abs-2103-16710}.
Here we use a similar concept in Monotonic Chunk-wise Attention~\cite{DBLP:conf/iclr/ChiuR18}, which is already widely used for speech recognition~\cite{DBLP:journals/corr/abs-2005-00205}.
However, we directly use the attention weights as a transition criterion without modifications to the training process.
We introduce our method in detail in Section~\ref{ssec:inf}.
Non-attentive Tacotron~\cite{DBLP:journals/corr/abs-2010-04301} approaches the alignment issue by replacing the attention mechanism of autoregressive models with an external duration predictor.
We will show that this technique actually degrades the performance of MQTTS.

\textbf{Non-autoregressive TTS.}
The focus of the TTS community gradually shifted toward non-autoregressive models~\cite{ pmlr-v139-kim21f,NEURIPS2020_5c3b99e8,NEURIPS2019_f63f65b5} due to their better robustness, inference speed, and lower training difficulty.
Recently, they have been shown to surpass autoregressive models in terms of naturalness on benchmark corpora.
Typically, a duration predictor is trained to predict the text-audio alignment, and a flow-based network is trained to synthesize given the alignment, phoneme sequence, and sampling Gaussian noise.
VITS~\cite{pmlr-v139-kim21f}, a recent end-to-end model that uses a stochastic duration predictor for better diversity, achieves near-human naturalness on the VCTK~\cite{VCTK} dataset.
While there are many non-autoregressive TTS systems, we choose VITS as the representative of this line of research and analyze its performance on real-world data compared to autoregressive models.

\textbf{Quantization-based Generative Models.}
An emerging line of research~\cite{pmlr-v139-ramesh21a,Esser_2021_CVPR} has shown success in image synthesis by adopting a two-stage approach.
A quantizer is trained to encode the image into discrete tokens, then an autoregressive transformer is used to model the sequential distribution of the image tokens.
DiscreTalk~\cite{DBLP:journals/corr/abs-2005-05525} adopts a similar framework for TTS, discretizing speech using a VQVAE and a transformer for autoregressive generation.
Unit-based language models~\cite{10.1162/tacl_a_00430} (uLMs) train the transformer on discrete tokens derived from representations of self-supervised (SSL) models, and a unit-to-speech (u2s) vocoder is separately trained to map the tokens to speech.
VQTTS~\cite{du22b_interspeech} further leverages the u2s vocoder for TTS by training a non-autoregressive model mapping text to discrete tokens.
MQTTS differs from these works in the use of multi-code, and the architecture designed for multi-code generation and monotonic alignment.
MQTTS also learns codebooks specialized for speech synthesis instead of using SSL representations which may not contain sufficient information for speech reconstruction.
%In addition, we use audio prompts for noise reduction and verify their importance for real-world speech synthesis in Section~\ref{sec:res}.

\textbf{TTS on real-world speech.}
An approach to using real-world speech for TTS is to incorporate denoising.
Recent work~\cite{DBLP:conf/icassp/Zhang00LZQZL21} trained the denoising module and the TTS system jointly with external noise data.
Our work focuses on the TTS system itself instead of explicit noise modeling, and we utilize a simple audio prompt for noise reduction without additional training.
There is also research~\cite{DBLP:journals/corr/abs-2107-02530} focusing on modeling characteristics of spontaneous speech, such as filled pauses (i.e., \textit{um}, \textit{uh}).
We do not design specific architectures for these, but rather let the models learn implicitly.

\begin{figure*}[!htb]
    \centering
    \includegraphics[width=0.8\textwidth]{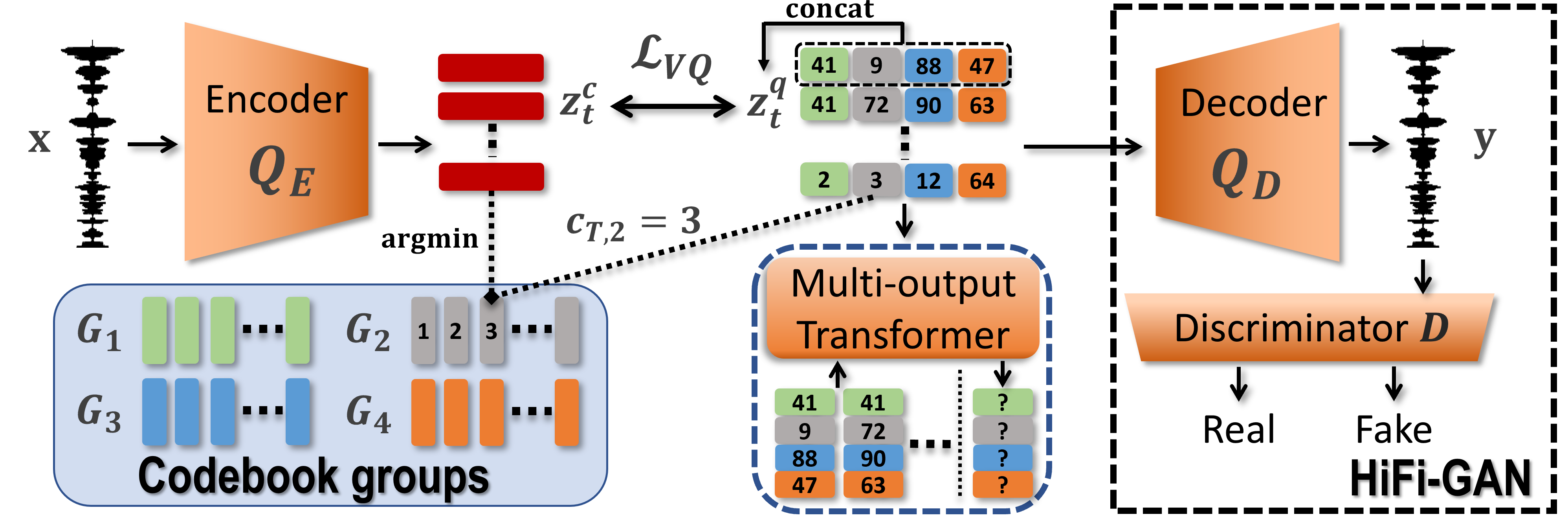}
    \caption{Overview of MQTTS. We use different colors to represent the 4 codes from distinct codebooks $G_i$.}
    \label{fig:VQmodule}
\end{figure*}
\section{MQTTS}
\label{sec:model}
We now introduce the model architecture as well as the training and inference procedures of MQTTS.
The training of our model has two stages.
In the first stage, as shown in Figure~\ref{fig:VQmodule}, a quantizer is trained to map the raw waveform $\vb{x}$ into discrete codes $\vb{c}=\{c_t\}$, $t\in [1, \cdots, T]$ by jointly minimizing a quantization loss $\mathcal{L}_{VQ}$ and the reconstruction loss between the input $\vb{x}$ and output $\vb{y}$.
A discriminator  guides the reconstruction using an additional adversarial loss.
We denote the pair of encoder and decoder in quantizer as $Q_E$, $Q_D$, the discriminator as $D$, and the learnable codebook embeddings as $G_i$ where $i\in [1, \cdots, N]$ for $N$ codebooks.
In the second stage, we fix the quantizer and train a transformer to perform autoregressive generation on $\vb{c}$, conditioned on a speaker embedding $s$ to enable multi-speaker TTS and a phoneme sequence $\vb{h}$.
Here we use bold math symbols to denote sequential representations across time.

\subsection{Quantization of Raw Speech}
\label{ssec:quant}
\textbf{Quantizer Architecture.}
We adopt HiFi-GAN~\cite{10.5555/3495724.3497152} as the backbone architecture for quantizer decoder $Q_D$ and the discriminator $D$, as its structure has been shown to produce high-quality speech given the mel-spectrogram.
We replace the input mel-spectrogram with the learned embedding of the discrete codes.
As for $Q_E$, we reverse the architecture of HiFi-GAN generator by substituting deconvolution layers with convolution layers.
We observed that naively adopting the HiFi-GAN architecture leads to the explosion of quantization loss $\mathcal{L}_{VQ}$ and training divergence.
We attribute this to the unnormalized output of residual blocks in HiFi-GAN.
Therefore, we apply group normalization~\cite{DBLP:journals/ijcv/WuH20} with 16 channels per group on the output of all residual blocks before aggregation.
In addition, we broadcast-add the speaker embedding $s$ on the input to the decoder $Q_D$.

\textbf{Multiple Codebook Learning.}
In Section~\ref{ssec:quantana}, we will show empirically that a single codebook results in poor reconstruction of the input signal for real-world speech.
We thus leverage the concept in vq-wav2vec~\cite{DBLP:conf/iclr/BaevskiSA20} to realize quantization with multiple codebooks.
As shown in Figure~\ref{fig:VQmodule}, each $c_t$ is now represented by $N=4$ codes, and the final embedding is the concatenation of the embeddings of the $N$ codes.
We denote that the individual code corresponding to codebook $G_i$ as $c_{t,i}$.
To get the discrete code $c_{t,i}$ from $z^c_t$, which is the output of $Q_E$, we slice the embedding dimension of $z^c_t$ into $N$ equal parts, denoting each part as $z^c_{t, i}$:
\begin{gather}
    c_{t,i} = \argmin_{z\in G_i} || z^c_{t,i} - z ||_2, 
\end{gather}
Then $z^q_{t}$, the input to the decoder is obtained by the straight-through gradient estimator:
\begin{gather}
    z^q_{t,i} =  z^c_{t,i} + \text{sg}[G_i(c_{t,i}) - z^c_{t,i}], \text{ } z^q_t = \text{concat}[z^q_{t,i}]
\end{gather}
where sg[$\cdot$] is the stop gradient operator, $G_i(\cdot)$ returns the embedding of the given code, and the concat[$\cdot$] operator concatenates all codes on the embedding dimension.
The final loss $\mathcal{L}_F$ is a combination of the HiFi-GAN loss term $\mathcal{L}_{GAN}$ and an additional quantization loss $\mathcal{L}_{VQ}$:
\begin{gather}
\label{eq:l_f}
    \mathcal{L}_F = \lambda\mathcal{L}_{VQ}(Q_E, Q_D, G) + \mathcal{L}_{GAN}(Q_E, Q_D, G, D)
\end{gather}
\begin{align}
\label{eq:l_vq}
\begin{split}
    \mathcal{L}_{VQ}(Q_E, Q_D, G) = &\mathbb{E}_x\biggl[\frac{1}{T}\sum_{i=1}^{N}\sum_{t=1}^{T} (|| \text{sg}[z^c_{t,i}] - z^q_{t,i} ||^2_2\\
    &+ \gamma||z^c_{t,i} - \text{sg}[z^q_{t,i}] ||^2_2)\biggr]
\end{split}
\end{align}
where $\gamma$ and $\lambda$ are hyper-parameters for the commitment loss weight and quantization loss weight.
We leave the detailed definition of $\mathcal{L}_{GAN}$ in the HiFi-GAN paper.
Note that our $\mathcal{L}_{VQ}$ is different from previous work in image generation; we do not include the reconstruction loss as the mel-spectrogram loss in $\mathcal{L}_{GAN}$ has already served this purpose.

\begin{figure}[!htb]
\centering
    \includegraphics[width=0.95\columnwidth]{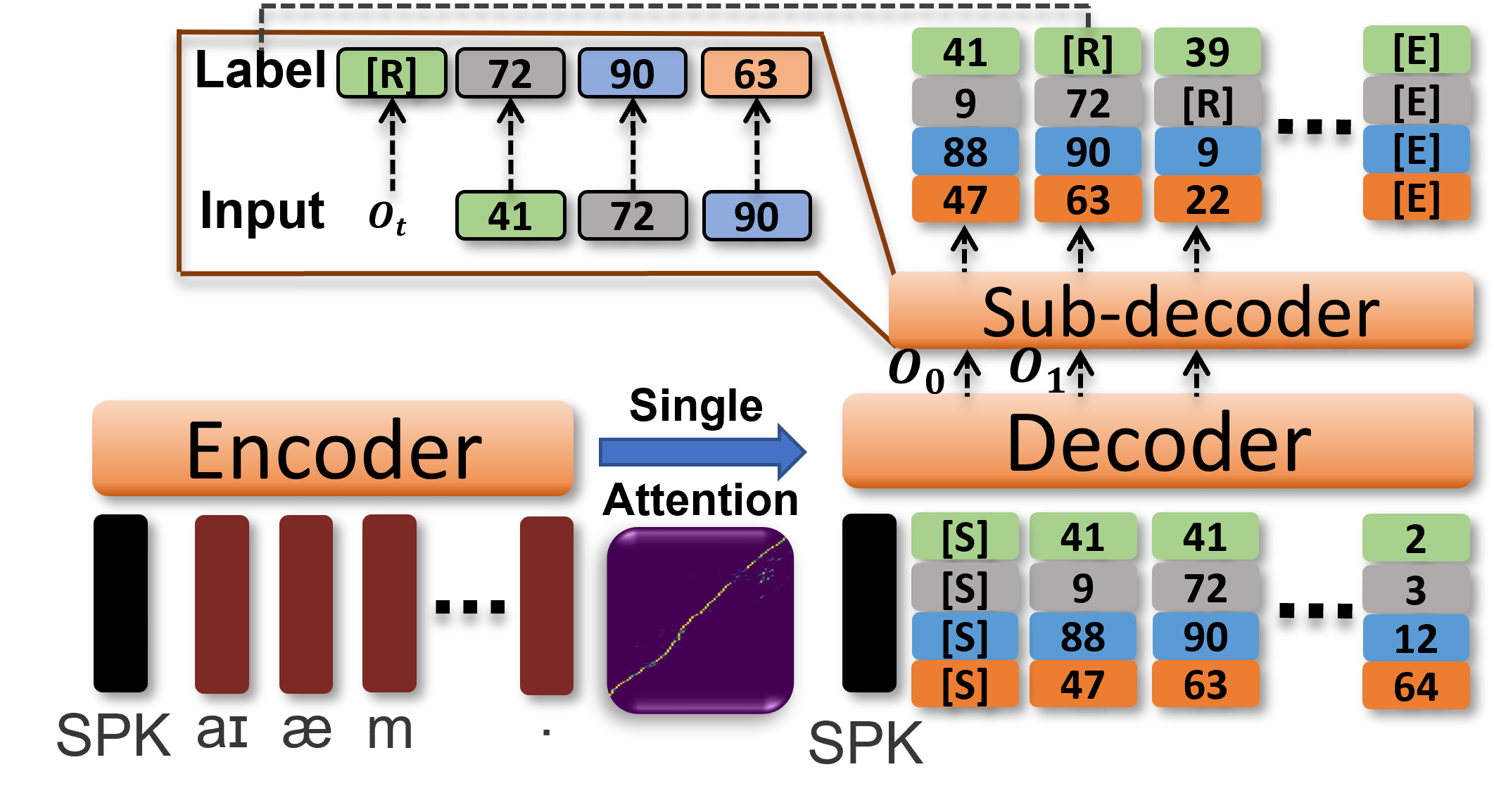}
\caption{Detailed view of the multi-output transformer during training. [R] refers to the repetition token. [S] and [E] are the start and end tokens respectively. SPK refers to the processed speaker embedding.}
\label{fig:tts}
\end{figure}

\subsection{Conditional Synthesis with Transformer}
In the second stage, we train a transformer to autoregressively generate $c_t$ given the past sequence $\vb{c}_{<t}$.
As presented in Figure~\ref{fig:tts}, the transformer is additionally conditioned on a global speaker embedding $s = E_{SPK}(\vb{x})$ and the phoneme sequence $\vb{h}$.
We use a pretrained speaker embedding extractor\footnote{https://huggingface.co/pyannote/embedding} followed by two linear layers with ReLU activation for $E_{SPK}$.
The transformer is trained to maximize log-likelihood:
\begin{gather}
    \mathcal{L}_T = \mathbb{E}_{\vb{c}}\left[-\sum_{t}\log{p(c_{t}\mid \vb{c}_{<t}, s, \vb{h})}\right]
\end{gather}

\textbf{Architectural Design.}
Different from the commonly used encoder-decoder architecture which contains cross attention from all decoder layers with the output encoder state, we adopt a different configuration.
The typical encoder-decoder architecture assumes that the two inputs have complex semantic correlations that need modeling from low-level to high-level representations, which is true for many NLP tasks.
However, in TTS, the cross-attention is simply a unique alignment map that benefits less from pre-estimating the alignment pattern using low-level audio representations.
As a result, we apply a single cross-attention layer only on the last layer of the decoder and use single-head cross-attention for the unique alignment.
More importantly, having a unique alignment map enables the use of monotonic alignment which we will introduce in Section~\ref{ssec:inf}.
We use ALiBi~\cite{alibi} to replace positional encoding to enable our model to extrapolate to long input sequences and syntheses for both encoder and decoder.

\textbf{Multi-output Transformer.}
While sampling techniques~\cite{Holtzman2020The,fan-etal-2018-hierarchical} play an important role in the performance of autoregressive models, it is not straightforward to adopt these to our multi-code setting.
If we simply use a linear layer to predict $N$ codes at a time, we can only access the distribution of each code individually.
Nucleus sampling~\cite{Holtzman2020The} will be theoretically infeasible, as it requires top-p candidates from the joint distribution of $N$ codes.
To address this, we use an additional transformer decoder to explicitly model the conditional distribution of codes, and name it the Sub-decoder module.
Different from the main decoder operating across the time sequence, Sub-decoder operates sequentially over the codebook groups of the fixed size $N$.
As shown in Figure~\ref{fig:tts}, at time $t$, the Sub-decoder accepts $O_{t}$, the output state of the main decoder, as starting condition and sequentially generates $c_{t,i}$ given $\vb{c}_{t,<i}$.
We use separate embeddings for each codebook group without positional encoding.
This configuration enables us to apply nucleus sampling at time $t$ on each of the conditional distributions $p(c_{t,i}\mid \vb{c}_{t,<i}, O_{t})$ respectively.
After generating the whole $c_t$, $c_t$ is fed back autoregressively to the main decoder for the generation of the next step $O_{t+1}$.

\textbf{Repetition Token.}
Speech signals often show high temporal similarity in their representations, making the transformer predict repeating codes consecutively.
We model such transitions explicitly by introducing an additional token, a repetition token.
Consecutive repeated codes are replaced by this token, and decoded back to the original token in inference.

\subsection{Inference}
\label{ssec:inf}
For the transformer inference, we adopt nucleus sampling~\cite{Holtzman2020The}, and the fully decoded codes are passed to $Q_D$ to reconstruct the raw waveform.

\textbf{Monotonic Alignment.}
We denote $A(i, j)$ as the cross attention value between encoder state at step $i$ and decoder state at step $j$.
Since our cross attention is single-head and single-layer, $A(i, j)$ is unique.
Then during inference, given a new decoder state at time $t$, we only calculate its cross attention with $N_w$ encoder states at $[b_k, b_k+1,\cdots,b_k+N_w]$ instead of the entire encoder state sequence, and $b_k$ is defined recursively by:
\begin{gather}
    b_0 = 0, b_{k+1} = \begin{cases}
    b_k + 1,& \text{if } \frac{\exp{A(b_k, k)}}{\sum_{i=0}^{N_w}\exp{A(b_k, k+i)}} < \frac{1}{N_w}\\
    b_k,              & \text{otherwise}
    \end{cases}
\end{gather}
Intuitively, the decoder can attend only to a certain context window $N_w$ of encoder states at a time sequentially, and the context window steps forward only if the Softmax attention weight of the first encoder state (at $b_k$) is lower than a given threshold (here we set it to ${N_w}^{-1}$).
In general, smaller $N_w$ leads to stricter monotonic alignment at the cost of a smaller phonetic context.
Practically, we found $N_w$ ranging from 3 to 6 works well judging from qualitative listening.
Such construction also enables us to use the alignment progress as a stopping criterion instead of predicting the stop token which is vulnerable to overfitting on the training utterance duration.

\textbf{Audio Prompt.}
When trained on a noisy corpus, it seems inevitable that undesired background noise will be sampled during inference.
Here we use a simple trick to prompt the model to synthesize clean speech.
We prepend 3 frames of codes that are encoded by $Q_{E}$ from a silence clip before the actual synthesis.
Clean speech is often preceded by clean silence, we expect it to encourage the model to continue the synthesis without background noise.

\section{Experimental Setup}
\label{sec:exp-stepup}
\subsection{Datasets}
\label{ssec:data}
We train our model on GigaSpeech~\cite{GigaSpeech2021}, an ASR corpus containing transcribed audio from audiobooks, Podcasts, and YouTube with a 16 kHz sampling rate.
We use only the Podcast and YouTube speech and apply a data-cleaning pipeline detailed in Appendix D to remove ill-conditioned speech.
The resulting dataset contains 896 hours for training and 2.4 hours for validation with an utterance duration ranging from 5 to 15 seconds.
For human evaluation, to ensure speaker diversity, we select 40 utterances from different speakers using the test set of VoxCeleb~\cite{Nagrani19} as the speaker reference audios, which are also noisy spontaneous speech.
We randomly select the corresponding text from GigaSpeech and remove those transcriptions from training and validation sets.

\subsection{Training and Model Configuration}
We use $\gamma = 0.25$ in Equation~\ref{eq:l_vq} and $\lambda = 10$ in Equation~\ref{eq:l_f} to balance the relative high $\mathcal{L}_{GAN}$.
For training the quantizer, Adam optimizer~\cite{kingma2014adam} is used with $\beta_1 = 0.5$, $\beta_2 = 0.9$.
%, a $2\times 10^{-4}$ initial learning rate and $0.999$ learning rate decay per epoch.
For the transformer, we use Adam optimizer with $\beta_1 = 0.9$, $\beta_2 = 0.98$ with a batch size of $200$ input frames per GPU.
The learning rate is linearly decayed from $2\times 10^{-4}$ to $0$.
All models are trained using 4 RTX A6000 GPUs in bfloat16 precision.
We trained the quantizer for 600k steps and the transformer for 800k steps.
We leave optimization details in the released code.
% where we discover little performance improvement.
%The quantizer training took about 3 days, and the transformer training took about 5 days.
For inference, we use nucleus sampling with $p=0.8$.

\textbf{Baseline Models.}
We use 6 layers for both encoder and decoder for Transformer TTS~\cite{Li_Liu_Liu_Zhao_Liu_2019} and add the speaker embedding with the same approach as our MQTTS.
We train a HiFi-GAN vocoder on GigaSpeech with the official implementation\footnote{https://github.com/jik876/hifi-gan} as its vocoder.
We also trained Tacotron 2~\cite{8461368} using the implementation from NVIDIA\footnote{https://github.com/NVIDIA/DeepLearningExamples}, and broadcast-added the speaker embedding to the encoder state following GST Tacotron~\cite{Wang2018StyleTU}.
For VITS, we modified from the official implementation and changed their fixed speaker embedding into our own speaker embedding module, and trained it for 800k steps.
We follow the original paper to use $0.8$ for the noise scaling of the stochastic duration predictor and $0.667$ for the prior distribution.

\textbf{Model Variants.}
To test the efficacy of the components in MQTTS, we evaluate two ablated versions of the model: one without monotonic alignment, and the other with the sub-decoder module replaced by the same number of linear layers.
Nucleus sampling is applied to each code independently for the version without Sub-decoder.
We also evaluate the scalability of model parameters with performance.
We report three versions of models: 40M, 100M, and 200M, which differ in the size of transformers while using the same quantizer.
The context window $N_w$ is set to 4 for 40M, 100M, and 3 for 200M version due to the clearer alignment.
The detailed architecture of each version is in Appendix A.
To better compare with MQTTS, we also trained a single cross-attention version of Transformer TTS with unique alignment.
Finally, we evaluate the approach from Non-attentive Tacotron~\cite{DBLP:journals/corr/abs-2010-04301} on improving alignment robustness.
We replace the attention alignment of Transformer TTS (single cross-attention version) and MQTTS with that produced by the duration predictor from VITS.

\subsection{Evaluation Metrics}

\textbf{ASR Error Rate.}
Error rate from Automatic Speech Recognition (ASR) models is a common metric to assess synthesis intelligibility~\cite{DBLP:conf/icassp/HayashiYIY0TTZT20}.
First, we adopt relative character error rate (RCER) to evaluate our quantizer model.
By relative, we mean using the transcription of the input utterance by the ASR model as the ground truth.
We believe that this better reflects how the reconstruction affects intelligibility.
We adopt ASR model\footnote{https://zenodo.org/record/4630406\#.YoT0Ji-B1QI} pretrained on GigaSpeech from ESPNet~\cite{watanabe2018espnet} for this purpose.
We also report the typical word error rate (WER) for the whole TTS system as an objective metric of intelligibility.
We use the video model of Google Speech-to-text API as the ASR model and evaluate 1472 syntheses.

\textbf{Human Evaluation.}
We report two 5-scale MOS scores, MOS-Q and MOS-N for quantizer evaluation, and only MOS-N for comparing TTS systems.
MOS-Q asks the human evaluators to score the general audio quality of the speech, while MOS-N evaluates the speech's naturalness.
We used the Amazon Mechanical Turk (MTurk) platform.
Details of the human experiments are in Appendix B.
We show both the MOS score and the 95\% confidence interval.

\textbf{Prosody FID Score (P-FID).}
For the evaluation of prosody diversity and naturalness, we visit the Fréchet Inception Distance (FID)~\cite{NIPS2017_8a1d6947}, a widely-used objective metric for the evaluation of image generative models.
FID calculates the 2-Wasserstein distance between real and synthesized distributions assuming they are both Gaussian, reflecting both naturalness and diversity.
A pretrained classifier is used to extract the dense representation for each sample.
In TTS, FID is seldom used due to the insufficient dataset size.
However, the 500k utterances in GigaSpeech afford the sampling complexity of FID.
We use the dimensional emotion classifier released by \cite{https://doi.org/10.48550/arxiv.2203.07378} pretrained on MSP-Podcast~\cite{Lotfian_2019_3}.
We use the input before the final decision layer and scale it by 10 to calculate the FID score.
We randomly sample 50k utterances from the training set and generate 50k syntheses using the same set of text but shuffled speaker references.

\textbf{Speaker Similarity Score (SSS).}
In multi-speaker TTS, it is also important to evaluate how well the synthesizer transfers speaker characteristics.
Here we use the cosine similarity between the speaker embedding of the reference speech and that of the synthesized speech.
We evaluate the same 50k samples used for P-FID.

\textbf{Mel-cepstral Distortion (MCD).}
We follow previous work~\cite{DBLP:journals/corr/abs-2110-07840} and use MCD to measure the reconstruction error between ground truth and synthesis, given the same text and speaker reference speech.
We extract 23 mel-cepstral coefficients (MCEPs) and use Euclidean distance as the metric.
Dynamic Time Wraping~\cite{dtw} is applied for the alignment.
We calculate MCD using the ground truth utterances from the validation split.

\section{Results}
\label{sec:res}
\subsection{Quantizer Analysis}
\label{ssec:quantana}
\begin{table}[!tb]
    \centering
    \caption{Comparison of quantizer and vocoder reconstruction quality on VoxCeleb test set. HF-GAN is HiFi-GAN.}
    \label{tab:vocoder:Giga}
    \begin{tabular}{l c c c c c}
        \toprule
        Method & \begin{tabular}{@{}c@{}}Code size \\ (\#groups)\end{tabular} & \begin{tabular}{@{}c@{}}RCER \\ (\%) \end{tabular} & \begin{tabular}{@{}c@{}}MOS-Q \\ (95\% CI) \end{tabular}  & \begin{tabular}{@{}c@{}}MOS-N \\ (95\% CI) \end{tabular} \\
        \midrule
        \textit{GT} & \textit{n/a} & \textit{n/a} & $3.66(.06)$ & $3.81(.05)$ \\
        \midrule
        \textit{HF-GAN} & \textit{n/a} & \textbf{12.8} & $3.47(.06)$ & $3.62(.06)$ \\
        \midrule
        \multirow{4}{*}{\begin{tabular}{@{}c@{}}\textit{MQTTS} \\ \textit{Quant.}\end{tabular}} & 1024 (1) & 56.5 & $3.38(.07)$ & $3.49(.06)$ \\
         & 65536 (1) & 59.9 & $3.40(.06)$ & $3.48(.06)$ \\
         & 160 (4) & 19.7 & $\textbf{3.63}(.06)$ & $3.67(.06)$ \\
         & 160 (8) & 14.2 & $3.56(.06)$ & $\textbf{3.71}(.06)$ \\
        \bottomrule
    \end{tabular}
\end{table}
Table~\ref{tab:vocoder:Giga} shows the reconstruction quality of the quantizer.
From RCER, it is clear that using a single codebook ($N=1$) drastically distorts the input phone sequence.
Their distinctive low MOS-Q and MOS-N scores also suggest that $N=1$ is not sufficient to generate high-quality and natural speech.
Even if we span a much larger codebook size, 65536, there is almost no difference compared to the 1024 codes across all metrics.
This result indicates the necessity of having multiple code groups, as a single codebook fails even to reconstruct the original input, not to mention when applied to the whole pipeline.
Interestingly, we find that a single codebook is sufficient for LJSpeech~\cite{ljspeech17}, suggesting multiple codebooks are required for the complexity of multi-speaker real-world speech.
When compared to the mel-spectrogram based HiFi-GAN (HF-GAN in Table~\ref{tab:vocoder:Giga}), our quantizer is slightly higher in terms of RCER but distinctively better in speech quality (MOS-Q).
For the remainder of our experiments, we use $N=4$ with a 160 code size, as we observe little perceptible quality and intelligibility difference when we increase from $N=4$ to $8$.

\begin{figure*}[!htb]
    \centering
    \includegraphics[width=\textwidth]{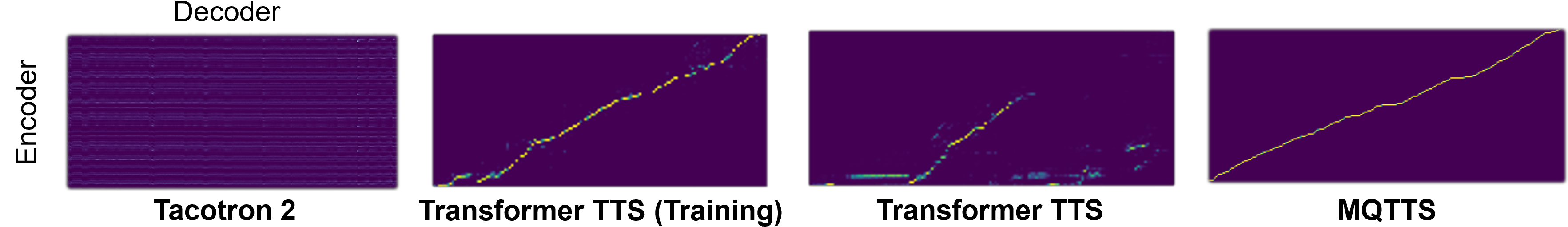}
    \caption{Comparison of inference encoder-decoder alignment of different models. For Transformer TTS we picked one of the cross-attentions which learn alignment. See Appendix C for the cross-attentions of all heads and layers.}
    \label{fig:alignment}
\end{figure*}
\begin{table*}[!htb]
    \centering
    \caption{Comparison of TTS models. MOS is with 95\% confidence interval. MCD is with one standard deviation.} %We only evaluate MOS scores that are informative for comparison.}
    \label{tab:architecture:Giga}
    \begin{tabular}{l|c|c c c c|c}
        \toprule
        Method & \#Params. & WER (\%) $\downarrow$ & SSS $\uparrow$ & P-FID $\downarrow$ & MCD $\downarrow$ & MOS-N $\uparrow$ \\
        \midrule
        \textit{GT} & \textit{n/a} & 14.9 & \textit{n/a} & 0.11 & \textit{n/a} & 4.01$\pm$0.06 \\
        \midrule
        \multirow{2}{*}{\textit{VITS}} & $\approx 40$M & 28.4 & 0.718 & 15.48 & 8.94$\pm$1.16 & 3.84$\pm$0.06\\
         & $\approx 100$M & 24.8 & 0.734 & 12.04 & \textbf{8.73}$\pm$1.19 & 3.84$\pm$0.06\\
        \midrule
        \textit{Transformer TTS} & \multirow{3}{*}{$\approx 100$M} & 82.6 & 0.434 & 442.29 & 12.49$\pm$2.77 & 2.34$\pm$0.09\\
        \textit{Transformer TTS (w. single-attention)} & & 90.2 & 0.331 & 491.49 & 12.79$\pm$4.35 & 2.23$\pm$0.10\\
        \textit{Transformer TTS (w. duration predictor)} & & 74.3 & 0.451 & 196.09 & 11.70$\pm$2.12 & 2.75$\pm$0.09\\
        \midrule
        \multirow{3}{*}{\textit{MQTTS}} & $\approx 40$M & 24.9 & 0.727 & 11.67 & 10.67$\pm$1.78 & 3.79$\pm$0.06\\
         & $\approx 100$M & 22.3 & 0.751 & 8.58 & 10.22$\pm$1.68 & 3.87$\pm$0.06\\
         & $\approx 200$M & \textbf{22.2} & \textbf{0.762} & \textbf{6.21} & 10.17$\pm$1.70 & \textbf{3.89}$\pm$0.06\\
        \midrule
        \textit{MQTTS (w.o. monotonic alignment)} & \multirow{3}{*}{$\approx 100$M} & 34.3 & 0.740 & 10.20 & 10.86$\pm$2.00 & 3.76$\pm$0.06\\
        \textit{MQTTS (w.o. sub-decoder)} & & 27.0 & 0.740 & 12.12 & 10.47$\pm$1.81 & 3.75$\pm$0.07\\
        \textit{MQTTS (w. duration predictor)} & & 53.2 & 0.725 & 16.65 & 10.88$\pm$1.62 & 3.67$\pm$0.06\\
        \bottomrule
    \end{tabular}
\end{table*}
%\begin{table}[!htb]
%    \centering
%    \caption{Comparison of TTS models. MOS are with 95\% confidence interval.}
%    \label{tab:architecture:Giga}
%    \begin{tabular}{l|c c c c}
%        \toprule
%        Method & WER (\%) $\downarrow$ & MOS-N $\uparrow$ & MOS-S $\uparrow$ & MOS-R $\uparrow$ \\
%        \midrule
%        \textit{GT} & 10.9 & $3.99\pm 0.04$ & $3.23\pm 0.09$ & $4.04 \pm 0.03$ \\
%        \midrule
%        \textit{VITS} & $40.2$ & $3.92 \pm 0.04$ & $\mathbf{3.33} \pm 0.06$ & $3.99 \pm 0.04$ \\
%        \midrule
%        \textit{Transformer TTS} & $91.9$ & $3.05 \pm 0.08$ & - & -\\
%        \midrule
%        \textit{MQTTS} & $\mathbf{25.3}$ & $\mathbf{3.97}\pm 0.04$ & $3.25 \pm 0.06$ & $\mathbf{4.00} \pm 0.05$ \\
%        \bottomrule
%    \end{tabular}
%\end{table}
\subsection{Performance}
\label{ssec:performance}
The performance of TTS system training on the given real-world speech corpus are presented in Table~\ref{tab:architecture:Giga}.

\textbf{Autoregressive Models.}
Before comparing systems quantitatively, we showcase some qualitative properties of Tacotron 2 and Transformer TTS.
Figure~\ref{fig:alignment} compares autoregressive models on their alignments.
It shows that Tacotron 2 completely failed the inference alignment.
On the other hand, Transformer TTS can produce clear alignment during training.
For inference, however, only a sub-sequence of alignment is formed.
Perceptually, only several words at the beginning are synthesized for Transformer TTS, then it is followed by unrecognizable noise.
As for Tacotron 2, only high-frequency noise is generated.
We choose not to include Tacotron 2 for quantitative evaluations as its syntheses are not understandable.
From Table~\ref{tab:architecture:Giga}, it is clear that Transformer TTS is inferior to other systems in all measures.
The poor inference alignment contributes largely to the bad performance, including the distinctively high P-FID as the syntheses contain long segments of non-speech.
Applying the duration predictor resolves the alignment issue and significantly increases the performance, but there is still a sizable gap with other systems.
We also noticed that using single attention on Transformer TTS leads to worse performance, indicating having multiple cross-attentions may still increase the robustness despite the high similarity in their alignment patterns.
Additionally, the larger parameter size (additional cross-attention layers) may also be another factor.
Nonetheless, this comes at the cost of the inability to use external duration predictors or monotonic alignment, as some of the heads are trained to average over a larger context.

\textbf{Model Variants.}
We can observe the efficacy of monotonic alignment and Sub-decoder from Table~\ref{tab:architecture:Giga}.
Both ablated versions have lower MOS-N scores.
Without monotonic alignment, the model suffers from noticeably higher WER.
On the other hand, replacing Sub-decoder module impact less on intelligibility but leads to a lower P-FID score.
This is reasonable as monotonic alignment mainly mitigates repetition, deletion errors and has less influence on prosody, while applying Sub-decoder leads to better sampling which improves prosody naturalness and diversity.
However, these ablated versions of MQTTS are still much better than Transformer TTS, indicating the advantage of discrete codes over continuous mel-spectrogram on real-world speech.
Replacing the alignment with the external duration predictor drastically degrades performance.
Inspecting the samples reveals that phonations often transition unnaturally.
We believe that as the given alignment is much different from the planned alignment of the MQTTS, it forces abrupt transitions in the middle of phonation.

\begin{figure}[!htb]
\centering
    \subfloat[VITS.]{%
    \includegraphics[width=\columnwidth]{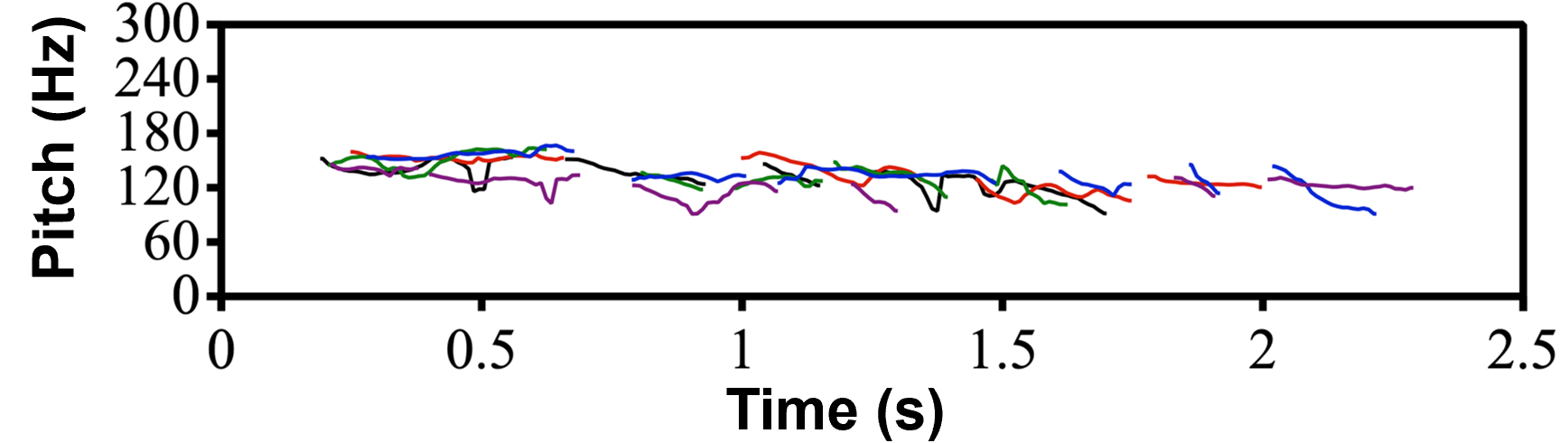}
    \label{fig:pitch_variation_vits}
    }
    
    \subfloat[MQTTS.]{%
    \includegraphics[width=\columnwidth]{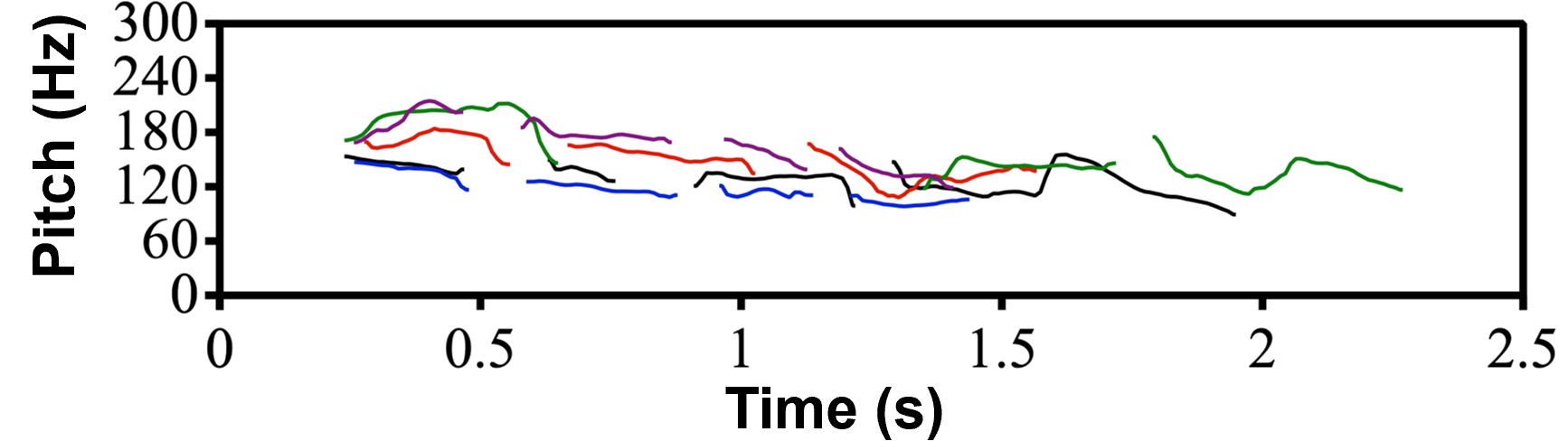}
    \label{fig:pitch_variation_vqtts}
    }
\caption{Pitch contour for the utterance: ``How much variation is there?'' from two models within the same speaker.}
\label{fig:pitch_variation}
\end{figure}

\textbf{Non-autoregressive Model.}
First, we notice that both VITS and MQTTS both have decent MOS-N scores.
However, the P-FID of MQTTS is noticeably lower than that of VITS, suggesting that MQTTS generates more diverse samples.
We conjecture that generating speech signals in parallel is a much harder task compared to autoregressive modeling.
Therefore, non-autoregressive models tend to focus on a smaller set of common speaking styles across the corpus due to insufficient capacity.
If we scale VITS from 40M to 100M, we see a decrease in P-FID with the same MOS-N, suggesting that bigger model capacity enables modeling of higher diversity, but without improvements in naturalness.
We did not include a 200M version of VITS as we failed to find a configuration that makes the training converge.
5 samples with the same text from both systems in Figure~\ref{fig:pitch_variation} further support our claim of the diversity comparison.
This might also explain the lower MCD of VITS, where the syntheses have conservative prosody patterns that are more tolerant in terms of MCD.
One possible example is the duration of internal pauses.
VITS syntheses mostly contain short intermittent pauses, while MQTTS often generates longer pauses while not uncommon in natural speech can potentially cause large misalignment with the ground truth, lowering MCD.

MQTTS performs better in terms of both intelligibility and speaker transferability.
We find MQTTS captures utterance-level properties (i.e., emotion, speaking rate) better compared to VITS.
For naturalness, we observe a consistently improving MOS-N of MQTTS as the model capacity grows.
It demonstrates different scaling properties: higher model capacity brings naturalness to MQTTS, but diversity to VITS.
Comparing the same parameter size (100M) for both VITS and MQTTS, MQTTS wins out in all metrics except MCD, which we explained earlier.
This suggests MQTTS is generally better than VITS, given enough resources.
Additionally, we observed overfitting for both 100M and 200M of MQTTS, but with a higher severity for the 200M version.
This explains the little improvement from 100M to 200M and suggests that a larger training corpus is needed for further improvement.

\textbf{Error Analysis.}
Despite the better average performance of MQTTS in Table~\ref{tab:architecture:Giga}, we find that it suffers from lower sampling robustness compared to non-autoregressive systems.
This is reasonable as higher diversity inevitably comes with a higher risk of sampling poor syntheses.
We observe unnaturally prolonged vowels in some samples with speaker reference speech of a slower speaking style, which is seldom the case for VITS.
In addition, samples that start with a poor recording environment often result in bad syntheses.
Deletion errors, which we consider more undesirable than substitution, are also more prevalent in MQTTS; it contributes for 8.4 out of 22.3\% WER in MQTTS-100M, but only 6.8 out of 24.8\% for VITS-100M.
We conjecture that as intermittent pauses are not annotated explicitly, and thus it encourages MQTTS to produce silence even if attending to a certain phoneme.
However, if the phones are successfully produced, they often sound more natural and intelligible than those in the syntheses of VITS.
We observe these errors gradually rectified as the model capacity increases (from 40M to 200M), suggesting that more data and larger models can eventually resolve these issues.

\begin{figure}[!htb]
    \centering
    \includegraphics[width=\columnwidth]{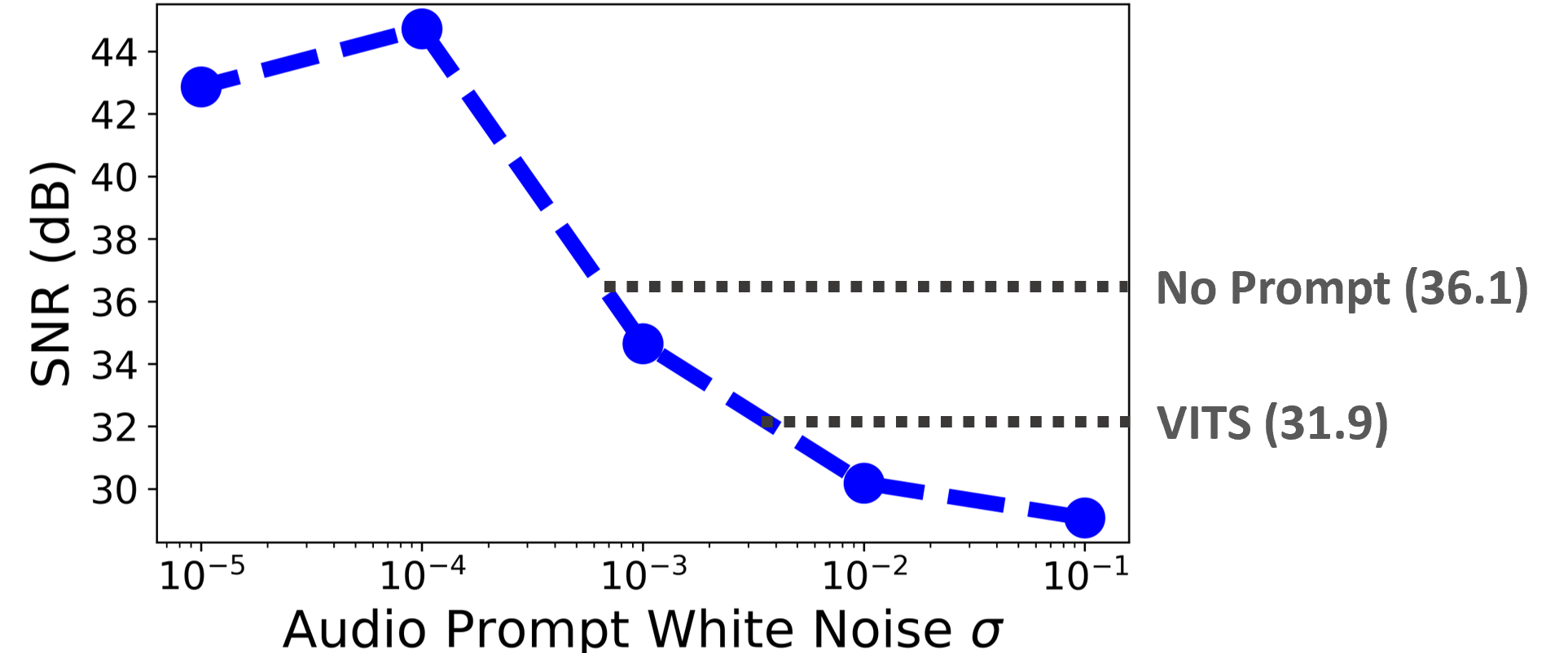}
    \caption{Comparison of SNR with different level of noise as audio prompt. SNR is calculated with 1472 syntheses.}
    \label{fig:snr}
\end{figure}

\subsection{Audio Prompt and SNR}
To better understand the effect of the audio prompts on the synthesis, we made the audio prompts white noise drawn from different standard deviation $\sigma$ (then encoded by $Q_E$).
Figure~\ref{fig:snr} presents the relationship between $\sigma$ and the signal-to-noise ratio (SNR).
We use the WADA-SNR algorithm~\cite{kim08e_interspeech} for SNR estimation.
From Figure~\ref{fig:snr}, it is clear that the SNR drops as $\sigma$ increases, confirming that the model is guided by the prompt.
Using $\sigma$ smaller than $10^{-4}$  effectively increases the SNR compared to not using any audio prompt.
All our other experiments are done using $\sigma=10^{-5}$.
We also noticed that VITS has a lower SNR.
Perceptually we can hear a small fuzzy noise universally across the syntheses.
We conjecture that VITS tries to also model and sample from the environmental noise, which is extremely difficult.
The unsuccessful modeling makes it synthesize only a single type of noise.
%Moreover, Figure~\ref{fig:snr} also shows that by adding a low energy white noise ($\sigma=1e-5$), we can further increase the SNR, achieving a much lower SNR than VITS.

\section{Conclusion}
\label{sec:conclusion}
On real-world speech, our empirical results indicate multiple discrete codes are preferable to mel-spectrograms for autoregressive synthesizers.
And with suitable modeling, MQTTS achieves better performance compared to the non-autoregressive synthesizer.
Nonetheless, a sizable gap still exists between our best-performing syntheses and human speech.
We believe that bridging this gap is crucial to the development of human-level communication for AI.
Acquiring more data is one straightforward solution.
In this regard, we are interested in combining and leveraging ASR models to transcribe real-world speech corpora.%, and analyzing the relationship between data size and performance.
On the other hand, better modeling can also be designed to mitigate the issues we mentioned in the error analysis.
For instance, silence detection can be used in the decoding process to prevent phoneme transitions before the phonation, mitigating deletion errors.
Additionally, we plan to further compare and incorporate self-supervised discrete speech and prosody representations with our learned codebooks.
%We believe that our method will be iteratively improved in the future by the research community.

\section*{Acknowledgments}
We are grateful to Amazon Alexa for the support of this research.
We thank Sid Dalmia and Soumi Maiti for the discussions and feedback.

\bibliography{main}

\end{document}